# Optimizing Bandwidth Sharing for Real-time Traffic in Wireless Networks

Sushi Anna George*, Vinay Joseph†

*Department of Electronics and Communication Engineering, National Institute of Technology Calicut,*
Kerala, India
Email: *sushiange@gmail.com, †vjoseph@nitc.ac.in,
ORCID: †0000-0001-6658-4406

*Abstract*—We consider the problem of enhancing the delivery of *real-time traffic* in wireless networks using *bandwidth sharing* between operators. A key characteristic of real-time traffic is that a packet has to be delivered within a delay deadline for it to be useful. The abundance of real-time traffic is evident in the popularity of applications like video and audio conferencing, which increased significantly during the COVID-19 period. We propose a sharing and scheduling policy which involves dynamically sharing a portion of one operator's bandwidth with another operator. We provide strong theoretical guarantees for the policy. We also evaluate its performance via extensive simulations, which show significant improvements of up to 90% in the ability to carry real-time traffic when using the policy. We also explore how the improvements from bandwidth sharing depend on the amount of sharing, and on additional traffic characteristics.

## I. Introduction

This paper considers wireless networks serving real-time traffic. A key characteristic of real-time traffic is that a packet *has to be delivered within a deadline*, failing which the packet becomes useless (or much less useful). Real-time traffic is key to several applications like video and audio conferencing, augmented and virtual reality applications, and online gaming. The relevance of real-time traffic increased substantially after the start of the COVID-19 pandemic. Video-conferencing generates a substantial volume of real-time traffic, and its market was predicted to grow by 12% between 2018-2023 [1]. The pandemic boosted this market, which is now expected to grow at the rate of 23% till 2027 [2].

To tackle the projected real-time traffic growth, we develop a solution based on *bandwidth sharing* to better support real-time traffic in wireless networks. Bandwidth sharing or resource pooling itself is not a novel concept (see more discussion in Section II). However, it is worthwhile to note here that bandwidth sharing is particularly attractive for real-time traffic, as the traffic has to be served within a tight timeline. This is evident from classic queuing theory results in [3] for $M/M/1$ queue with impatient customers having a deterministic deadline $D$. An impatient customer with deadline $D$ is similar to a packet of real-time traffic (which too has a deadline). From [3], the probability of successful service of a customer within deadline $D$ can be derived as:

$$P_{succ}(\mu, \rho, D) = \frac{1 - e^{\mu D(\rho - 1)}}{1 - \rho e^{\mu D(\rho - 1)}}.$$

Here $\mu$ is the average service rate of customers and $\rho$ is the traffic intensity. To assess the value of bandwidth sharing, we evaluate the percentage improvement in probability of successful service with sharing relative to that without sharing. Queue with sharing is obtained by pooling of arrivals and servers of two queues. Thus, it can be viewed as a new queue with double the arrival rate and service rate (i.e., $\mu$ is doubled whereas $\rho$ is unchanged). Hence, the percentage improvement in probability of successful service with sharing is

$$G_{pool}(\mu, \rho, D) = \frac{P_{succ}(2\mu, \rho, D) - P_{succ}(\mu, \rho, D)}{P_{succ}(\mu, \rho, D)}.$$

Fig. 1 below shows how the gain drops as deadline $D$ increases. The larger pool of resources (resulting from sharing) is better suited to address real-time traffic's urgency in demand for resources. Larger $D$ values can be viewed as traffic which is almost not real-time. A key takeaway here is that *bandwidth sharing is especially attractive for real-time traffic* (more than for non-real-time traffic), in particular due to the urgency in its demand for bandwidth.

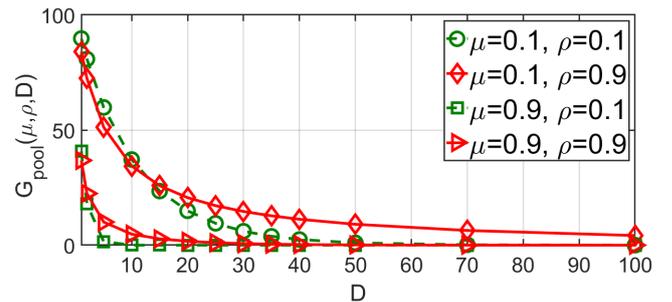

Fig. 1: Improvement in success probability with sharing is higher for shorter deadlines ($D$)

In this paper, we consider dynamic wireless bandwidth sharing between different operators across multiple geographical regions. The basic idea is illustrated in Fig. 2. The bandwidth sharing mechanism enables an operator $A$ with lower real-time traffic in region 1 to share some of its timeslots with another operator $B$ in the same region. Clearly, such bandwidth sharing requires strong incentives for the operators who are likely competitors competing for the same customers. Thus, we consider a sharing framework where the net amount of bandwidth shared between any two operators across all regions is forced to be equal. That is, even if operator $A$ shares more bandwidth with operator $B$ in region 1, the sharing framework requires that operator $B$ compensates for this by sharing more in other regions. The operators in this case may be using licensed spectrum, in which case the sharing can be realized using techniques like carrier aggregation [4]. The operators could alternatively be using unlicensed spectrum in a coordinated manner in colocated deployments (e.g., using NR-Unlicensed), where different operators coordinate to use different parts of the unlicensed spectrum (to avoid interference) for their regular operation and use sharing on top of it.

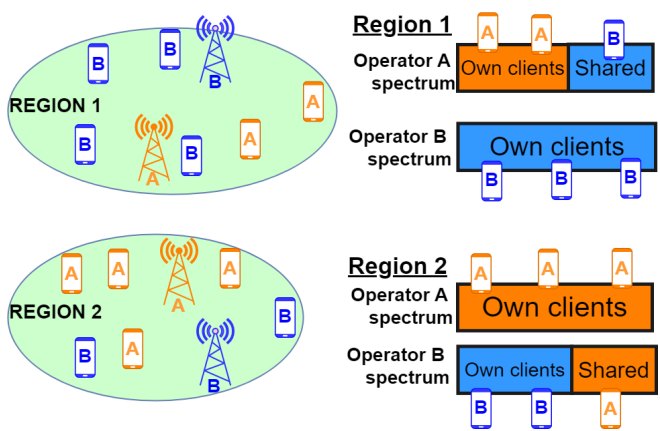

Fig. 2: Illustration of bandwidth sharing between operators across regions

Bandwidth sharing or spectrum sharing is being actively discussed in many forums across the world including government agencies, industry and other stakeholders. UK's telecom regulator Ofcom (see [5]) and Communications Technology Laboratory of the National Institute of Standards and Technology of the USA (see [6]) are studying spectrum sharing. Spectrum sharing is also an important area in the efforts (focusing on 3.5 GHz band) of Wireless Innovation Forum and Citizens Broadband Radio Service in USA (see [7]). India too has published a set of guidelines on spectrum sharing (see [8]). Despite the significant interest in spectrum sharing, there has been no work on spectrum sharing for real-time traffic. We address this gap in this paper.

### A. Main Contributions

Major contributions of this work are summarized below:

1) We develop a novel framework for bandwidth sharing for real-time traffic between multiple operators across multiple regions (see Section III).
2) We propose a joint sharing and scheduling policy, with strong optimality guarantees (see Section IV).
3) We provide insights into the nature of gains from bandwidth sharing using extensive simulations (see Section V).

### B. Organization of the paper

We discuss related work in Section II. System model is discussed in Section III. Section IV presents our sharing and scheduling policy and a related theoretical result. Simulation results are in Section V. We conclude in Section VI.

## II. RELATED WORK

**On bandwidth/spectrum sharing between operators:** [9] proposed protocols for cellular networks to redistribute excess call traffic on a spectrum band, to spectrum bands with excess capacity. [10]–[12] also propose spectrum sharing solutions and quantify the gains. Above papers do not however consider real-time traffic.

**On scheduling of real-time traffic in wireless networks:** There has been previous work on providing services for delay-constrained traffic in wireless networks. One of our key references is [13], which analyzed scheduling real-time traffic in unreliable wireless environments. [13] introduced a term *timely throughput* to measure the amount of real-time traffic that is successfully delivered. Further, [14] and [15] developed scheduling policies with optimality guarantees for real-time traffic for various scenarios (e.g., rate adaptation, time-varying channels). Scheduling real-time traffic with hard deadlines in a wireless ad hoc network by ensuring both timely throughput and data freshness guarantees for deadline-constrained traffic is considered in [16]. The system models in [13], [14], [15] and [16] rely on *frame-based* models for arrival and scheduling of real-time traffic. In frame-based models (which is used in this paper too), all traffic arrives at the beginning of a frame and has to be served by the end of the frame. The approach in [17] does not require this frame-based approach and uses an approach relying on randomization of the choice of transmitting links. Papers like [18], [19] have also approached this topic formulating the problem in terms of *age of information* and developed scheduling algorithms using deep reinforcement learning. Note that these previous works do not however explore solutions leveraging bandwidth sharing.

**On real-time traffic related to queuing theory:** Queuing theory studies related to customer impatience (and reneging after lapse of deadline) provide useful insights about serving of real-time traffic. By taking into account the M/M/c queue with independent exponentially distributed customer waiting times, Palm's groundbreaking study [20] examines queuing systems with impatient customers. [3] provides a closed form expression for loss probability of an $M/M/1$ with deterministic impatience distribution. Only a few papers like [21]–[23], however provide closed form expressions for useful metrics like probability of reneging. Real-Time Queuing Theory (RTQT) introduced by Lehoczky addresses the limitations of using classic queuing theory (which typically focuses on average behaviour) for real-time systems. RTQT was explained using an $M/M/1$ queue with Earliest Deadline First (EDF) approach [24]. More general settings were treated

TABLE I: Notation

| Notation | Description |
|---|---|
| $\mathcal{O}$ | Set of all operators |
| $i, j$ | Indices used for an operator |
| $\mathcal{R}$ | Set of all regions, $\mathcal{R} = \{1, 2, \ldots, R\}$ |
| $r$ | Index used for a region |
| $\mathcal{N}_r^i$ | Set of all users in the $r^{th}$ region of the $i^{th}$ operator |
| $n$ | Index used for a user |
| $k$ | Index of period |
| $T$ | Number of timeslots within a period |
| $\zeta^{(i,j)}$ | Bound on difference in sharing of any two operators |
| $A_{r,n}^i(k)$ | Number of packet arrivals of $n^{th}$ client in the $r^{th}$ region of the $i^{th}$ operator in the $k^{th}$ period |
| $A_r^i(k)$ | Number of packet arrivals in the $r^{th}$ region of the $i^{th}$ operator in the $k^{th}$ period |
| $b_{r,n}^i(k)$ | Scheduling indicator for $n^{th}$ client in $r^{th}$ region of $i^{th}$ operator, which takes value one when packet is scheduled in the $k^{th}$ period, otherwise zero. |
| $I_{r,n}^i(k)$ | Successful delivery indicator for $n^{th}$ client in $r^{th}$ region of $i^{th}$ operator, which takes value one when packet is successfully delivered in the $k^{th}$ period, otherwise zero. |
| $P_{r,n}^i$ | Probability of successful delivery of scheduled packet for $n^{th}$ client in $r^{th}$ region of $i^{th}$ operator. |
| $q_{r,n}^i$ | Minimum timely throughput requirement of the $n^{th}$ client in the $r^{th}$ region of the $i^{th}$ operator |
| $S_r^{j \to i}(k)$ | Number of timeslots shared by the $r^{th}$ region of the $j^{th}$ operator to the $r^{th}$ region of the $i^{th}$ operator in the $k^{th}$ period |
| $\sigma^{i \to j}(k)$ | Sharing debt owed by operator $i$ to operator $j$ upto the $k^{th}$ period |
| $\delta_{r,n}^i(k)$ | Delivery debt of the $n^{th}$ client in the $r^{th}$ region of the $i^{th}$ operator upto the $k^{th}$ period |

using RTQT in [25] and [26].

## III. SYSTEM MODEL

We consider the downlink of a wireless system spanning a set of regions $\mathcal{R}$ served by a set of operators $\mathcal{O}$. Let $\mathcal{N}_r^i$ denote the set of clients in region $r$ of operator $i$. We consider a slotted time model. We assume that each operator can transmit to at most one client in any timeslot in any given region. Further, we assume that an operator can transmit simultaneously in different regions without the transmissions interfering with each other. In short, there is no interference between transmissions of different operators in one region, different operators in different regions, or same operator across different regions (and there is no interference between transmissions to different clients of an operator in one region).

**Arrivals:** We model real-time traffic as in [15], where all the arrivals happen at the start of a collection of $T$ consecutive timeslots. Consecutive $T$ timeslots is referred to as a period. Let $A_{r,n}^i(k) \in \{0, 1\}$ denote whether a packet of client $n$ in region $r \in \mathcal{R}$ of operator $i \in \mathcal{O}$, arrives at the start of period $k$. Also, let $A_r^i(k)$ denote the total number of packet arrivals of all the clients in region $r \in \mathcal{R}$ of operator $i \in \mathcal{O}$ in period $k$. We model $\{A_{r,n}^i(k) : k \geq 1\}$ as a stationary irreducible Markov process with finite state space, and assume that they are independent for any two clients.

**Scheduling:** In one timeslot, we assume that at most one client can be scheduled by an operator in a region. Let $b_{r,n}^i(k)$ be the scheduling indicator for the $n^{th}$ client in $r^{th}$ region of $i^{th}$ operator. $b_{r,n}^i(k)$ takes value of one when a packet of client $n$ is scheduled in $k^{th}$ period, and is otherwise zero. We also have $b_{r,n}^i(k) \leq A_{r,n}^i(k)$, i.e., scheduling is considered only for a client with a packet arrival in period $k$. Note that at most $T$ clients can be scheduled in a period by an operator in a region, as there are only $T$ timeslots in a period.

**Wireless channel:** If a packet arrives for a client $n \in \mathcal{N}_r^i$ in a period $k$ and it is scheduled, we assume that it is delivered successfully with probability $P_{r,n}^i$. Note that this model allows capturing heterogeneous wireless channels, i.e., client, operator and region dependent channels. For instance, a client with good channel conditions can be modelled using a higher value for $P_{r,n}^i$, when compared to another client with poor channel conditions (e.g., at the edge of the coverage region). Let $I_{r,n}^i(k)$ be an indicator of successful delivery for the $n^{th}$ client in the $r^{th}$ region of the $i^{th}$ operator, which takes value one when packet is successfully delivered in the $k^{th}$ period, and is otherwise zero. We assume that $\{I_{r,n}^i(k) : k \geq 1\}$ are independent and identically distributed, and that they are independent for any two clients.

**Timely throughput requirement:** A packet that has arrived at the start of a period must be transmitted within the period (i.e., within $T$ timeslots) for its timely (or successful) delivery, or is dropped otherwise. A key metric thus for real-time traffic is the average number of timely or successful packet deliveries per period for a client, referred to as timely throughput of that client (as defined in [15]). Client $n \in \mathcal{N}_r^i$ requires timely throughput of at least $q_{r,n}^i$, which is expressed as the following probabilistic requirement for a small positive constant $\xi_1$:

$$\text{Prob}\left\{\frac{1}{K}\sum_{k=1}^{K} b_{r,n}^i(k) I_{r,n}^i(k) \geq q_{r,n}^i - \xi_1\right\} \to 1,$$
$$\text{as } K \to \infty, \ \forall n \in \mathcal{N}_r^i, \ \forall i \in \mathcal{O}, \ \forall r \in \mathcal{R}. \quad (1)$$

Note that the above requirement can be used to model packet loss requirements in video and audio conferencing. For instance, we can set $q_{r,n}^i$ as 0.95 times the packet arrival rate of a client, to match packet loss value of 5% mentioned in [27].

### A. Bandwidth Sharing Model

In any region, an operator can share one/more timeslots in a period with another operator. Note that this sharing is region-specific, and different sharing is possible in different regions. Let $S_r^{j \to i}(k)$ denote the number of timeslots of the $j^{th}$ operator shared with the $i^{th}$ operator in the $r^{th}$ region in the $k^{th}$ period. Here, $S_r^{i \to i}(k)$ is the number of timeslots of the $i^{th}$ operator used for its own clients in $r^{th}$ region in the $k^{th}$ period. Note that the bandwidth sharing considered here involves sharing the *entire* spectrum of an operator (in a region) over a few slots in every period. Also, note that sharing considered here is dynamic, requiring slot-by-slot coordination between operators.

Following is a key constraint in our sharing framework which bounds the difference in average number of timeslots shared by any pair of operators $i, j \in \mathcal{O}$:

$$\text{Prob}\left\{\frac{1}{K}\left|\sum_{k=1}^{K}\sum_{r\in\mathcal{R}}S_r^{j\to i}(k) - \sum_{k=1}^{K}\sum_{r\in\mathcal{R}}S_r^{i\to j}(k)\right| \leq \zeta^{(i,j)} + \xi_2\right\}$$
$$\to 1, \text{ as } K \to \infty, \quad (2)$$

where $\xi_2$ is a small positive constant. Here the parameter $\zeta^{(i,j)}$ controls the relative amount of sharing between operators $i$ and $j$. In particular, for small $\zeta^{(i,j)}$, (2) essentially ensures that each operator gives roughly as much it gets from another operator as part of sharing. Thus (2) incentivizes sharing (an operator receives only as much as it gives) and also limits over-sharing (an operator gives only as much as it receives).

Without sharing, the number of packets that can be scheduled by operator $i$ in region $r$ is bounded by $S_r^{i\to i}(k)$. With sharing, the number of packets that can be scheduled by operator $i$ in region $r$ is bounded by the sum of $S_r^{i\to i}(k)$ (timeslots owned by the operator) and $\sum_{j\in\mathcal{O}\setminus\{i\}}S_r^{j\to i}(k)$ (timeslots shared by other operators), i.e., we have:

$$\sum_{n\in\mathcal{N}_r^i}b_{r,n}^i(k) \leq \sum_{j\in\mathcal{O}}S_r^{j\to i}(k), \quad \forall i \in \mathcal{O}, \forall r \in \mathcal{R}, \forall k \in \mathcal{K} \quad (3)$$

A few additional constraints on sharing are given below:

$$S_r^{j\to i}(k) \geq 0, \quad \forall i,j \in \mathcal{O}, \forall r \in \mathcal{R}, \forall k \in \mathcal{K}; \quad (4)$$

$$\sum_{i\in\mathcal{O}}S_r^{j\to i}(k) \leq T, \quad \forall j \in \mathcal{O}, \forall r \in \mathcal{R}, \forall k \in \mathcal{K}; \quad (5)$$

$$S_r^{i\to i}(k) \leq A_r^i(k), \quad \forall i \in \mathcal{O}, \forall r \in \mathcal{R}, \forall k \in \mathcal{K}; \quad (6)$$

$$\sum_{i\in\mathcal{O}\setminus\{j\}}S_r^{j\to i}(k) \leq \max(T - A_r^j(k), 0), \quad (7)$$
$$\forall j \in \mathcal{O}, \forall r \in \mathcal{R}, \forall k \in \mathcal{K};$$

$$\sum_{j\in\mathcal{O}\setminus\{i\}}S_r^{j\to i}(k) \leq \max(A_r^i(k) - T, 0), \quad (8)$$
$$\forall i \in \mathcal{O}, \forall r \in \mathcal{R}, \forall k \in \mathcal{K}.$$

(4) requires that $S_r^{j\to i}(k)$ are non-negative. (5) captures that maximum number of timeslots available to an operator for its own use and for sharing with other operators is $T$. (6) specifies that number of timeslots that an operator can use for its own clients is limited by the total number of packet arrivals. (7) ensures that an operator shares timeslots with other operators only if there are timeslots left after scheduling own clients. (8) limits the number of timeslots received from other operators to the number of timeslots required after utilizing all $T$ timeslots.

*B. Feasibility Optimal Policy*

Observe from the preceding discussion that the key decision variables involved are the following:

- which client to schedule, i.e., $\forall k$ deciding $\mathbf{b}(k) = \left(b_{r,n}^i(k) : \forall n \in \mathcal{N}_r^i, \forall r \in \mathcal{R}, \forall i \in \mathcal{O}\right)$, and
- how much to share, i.e., $\forall k$ deciding $\mathbf{S}(k) = \left(S_r^{j\to i}(k) : \forall r \in \mathcal{R}, \forall i,j \in \mathcal{O}\right)$.

A *policy* specifies the above scheduling and sharing decision variables for each period $k$.

We let $\mathbf{q} = \left(q_{r,n}^i : n \in \mathcal{N}_r^i, \forall r \in \mathcal{R}, \forall i \in \mathcal{O}\right)$ and $\boldsymbol{\zeta} = \left(\zeta^{(i,j)} : i,j \in \mathcal{O}\right)$. Timely throughput requirement $\mathbf{q}$ and sharing bound $\boldsymbol{\zeta}$ is said to be *feasible*, if there exists a policy satisfying (1)-(8). Timely throughput requirement $\mathbf{q}$ and sharing bound $\boldsymbol{\zeta}$ is said to be *strictly feasible*, if there exists a constant $\kappa$ with $0 < \kappa < 1$, such that modified throughput requirement $\mathbf{q}/\kappa$ and sharing bound $\kappa\boldsymbol{\zeta}$ is feasible. A policy is said to be *feasibility optimal* if it satisfies every strictly feasible timely throughput requirement $\mathbf{q}$ and sharing bound $\boldsymbol{\zeta}$.

## IV. ONLINE FEASIBILITY-OPTIMAL SHARING AND SCHEDULING POLICY

Here, we present our sharing and scheduling policy, and discuss a related optimality result (Theorem 1). Designing a feasibility-optimal sharing and scheduling policy is not straightforward, especially due to the time averaging involved in the timely throughput constraint (1) and sharing constraint (2). To tackle this, we utilize 'debt parameters' or virtual queues $\delta_{r,n}^i(k)$ and $\sigma^{i\to j}(k)$, similar to those in [15], [28] etc. Here $\delta_{r,n}^i(k)$ is the virtual queue tracking delivery debt of the $n^{th}$ client in the $r^{th}$ region of the $i^{th}$ operator upto the $k^{th}$ period (similar to that in [15]). $\sigma^{i\to j}(k)$ denotes the sharing debt owed by operator $i$ to operator $j$ upto the $k^{th}$ period.

Our online policy basically minimizes the following function in each period $k$:

$$f(\mathbf{b}(k), \mathbf{S}(k), \boldsymbol{\delta}(k), \boldsymbol{\sigma}(k)) = -\sum_{r\in\mathcal{R}}\sum_{i\in\mathcal{O}}\sum_{n\in\mathcal{N}_r^i}\delta_{r,n}^i(k)b_{r,n}^i(k)P_{r,n}^i$$
$$+ \sum_{i\in\mathcal{O}}\sum_{j\in\mathcal{O}\setminus\{i\}}\sigma^{i\to j}(k)\left(\sum_{r\in\mathcal{R}}S_r^{j\to i}(k) - \sum_{r\in\mathcal{R}}S_r^{i\to j}(k)\right), \quad (9)$$

where $\boldsymbol{\delta}$ is a multi-dimensional array with entries $\delta_{r,n}^i$ for each $n \in \mathcal{N}_r^i$, $\forall r \in \mathcal{R}$, $\forall i \in \mathcal{O}$, and $\boldsymbol{\sigma}$ is a multi-dimensional array with entries $\sigma^{i\to j}$ for each $i,j \in \mathcal{O}$. Our policy is given below:

---
**Online Sharing and Scheduling Policy**

Scheduling $\mathbf{b}^*(k)$ and sharing $\mathbf{S}^*(k)$ for period $k$ are determined by solving the following optimization problem:

$$(\mathbf{b}^*(k), \mathbf{S}^*(k)) = \underset{(\mathbf{b}(k),\mathbf{S}(k))}{\operatorname{argmin}} \quad f(\mathbf{b}(k), \mathbf{S}(k), \boldsymbol{\delta}(k), \boldsymbol{\sigma}(k))$$
$$\text{s.t.} \quad (3), (4), (5), (6), (7) \text{ and } (8).$$

$\delta_{r,n}^i(k)$ and $\sigma^{i\to j}(k)$ are initialized to zero at $k = 1$, and updated in each period $k > 1$ as follows:

$$\delta_{r,n}^i(k+1) = \max(\delta_{r,n}^i(k) + q_{r,n}^i - b_{r,n}^{*i}(k)I_{r,n}^i(k), 0),$$
$$\forall n \in \mathcal{N}_r^i, i \in \mathcal{O}, r \in \mathcal{R}; \quad (11)$$

$$\sigma^{i\to j}(k+1) = \max(\sigma^{i\to j}(k) + \quad (12)$$
$$\sum_{r\in\mathcal{R}}S_r^{*,j\to i}(k) - \sum_{r\in\mathcal{R}}S_r^{*,i\to j}(k) - \zeta^{(i,j)}, 0),$$
$$\forall i \in \mathcal{O}, j \in \mathcal{O}\setminus\{i\}, k \in \mathcal{K}.$$

---

Observe that our policy does not explicitly include long term time-average requirements (1) and (2) as constraints. Rather, they are met by ensuring that the debts tracked using $\delta_{r,n}^i(k)$ and $\sigma^{i \to j}(k)$ are not too high. In particular, an under-scheduled client will tend to have a high value of $\delta_{r,n}^i(k)$ (see 11). This high value leads to a higher chance of scheduling of the client, since $\delta_{r,n}^i(k)$ scales $b_{r,n}^i(k)$ in the objective function (9). Similarly, an operator $i$ that has not shared much will tend to have a high value of $\sigma^{i \to j}(k)$ (see 12), and this leads to higher chance of sharing by the operator in subsequent periods, since $\sigma^{i \to j}(k)$ scales $S_r^{i \to j}(k)$ in the objective function (9).

### A. Optimality

Following theorem provides a strong theoretical performance guarantee for our joint sharing and scheduling policy. Its proof can be found in appendix A.

**Theorem 1.** *The sharing and scheduling policy $((\mathbf{b}^*(k), \mathbf{S}^*(k)) : k \geq 1)$ is feasibility optimal.*

Above result says that our sharing and scheduling policy is feasibility optimal (defined in Section III-B). Put another way, if there is any policy that meets a timely throughput requirement and sharing bound, then our policy too will meet them. The proof is omitted for brevity, and uses Lyapunov analysis (e.g., see [28]) using Lyapunov function below

$$\sum_{r \in \mathcal{R}} \sum_{i \in \mathcal{O}} \sum_{n \in \mathcal{N}_r^i} \left(\delta_{r,n}^i(k)\right)^2 + \sum_{i \in \mathcal{O}} \sum_{j \in \mathcal{O} \setminus \{i\}} \left(\sigma^{i \to j}(k)\right)^2.$$

## V. PERFORMANCE EVALUATION

In this section, we present simulation results for our sharing and scheduling policy. We primarily assess benefits of sharing by evaluating *percentage improvement in timely throughput* with sharing compared to that without sharing.

### A. Simulation settings

We consider two operators serving two regions (i.e., $\mathcal{R} = \{1, 2\}$) with 10 clients each. For roughly equal sharing between operators, we set $\zeta^{(i,j)}$ as a small value of 0.001. Let $\alpha_r^i$ denote the average number of packet arrivals for each client in region $r \in \mathcal{R}$ of operator $i \in \mathcal{O}$. We set $\alpha_1^1 = \alpha_2^2$ and $\alpha_2^1 = \alpha_1^2$, so that an operator will have a relatively low arrival rate in one region and a high arrival rate in the other. Except for Fig. 5 with variable $T$, we set length of a period $T = 5$ (the deadline of each packet too is five timeslots). We set $q_{r,n}^i = 0.95\alpha_r^i$ for each client, based on typical values like 5% for allowable packet loss (see [27]). We set $P_{r,n}^i = 0.99$ and $K = 10000$ periods.

### B. Simulation results

In Fig. 3, we study the impact of limiting the maximum sharing between operators. The figure shows percentage improvement in timely throughput for different limits. The limit on maximum sharing is imposed by capping the right hand sides of (7) and (8) with the limit. The results indicate that even modest sharing improves timely throughput, and that the improvement increases with more sharing. Further, the improvements are more when arrival rates of operators are imbalanced, i.e., when $|\alpha_1^1 - \alpha_1^2|$ is high. For many data points depicted in the figure, all clients meet their timely throughput targets with sharing, even when all are unable to meet them without sharing.

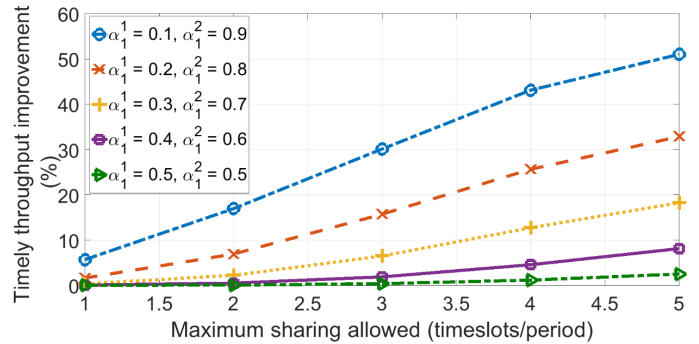

Fig. 3: Percentage improvement in timely throughput with limits on sharing

In Fig. 4a, we further zoom-in on the impact of imbalance in arrival rates $|\alpha_1^1 - \alpha_1^2|$. Here we set arrival rates in region 1 and 2 as scaled versions of $\beta_1$ and $\beta_2 = 1 - \beta_1$ respectively, where $\beta_1 \in \{0, 0.1, ..., 0.5\}$. Here too, the results indicate that higher the imbalance in arrival rates, higher are the timely throughput improvements with sharing. This is expected as unbalanced arrival rates allow more sharing, as the operator with low arrival rate in a region can give its resources to the other operator in that region, and the other operator reciprocates in the other region. This is also evident from Fig. 4b.

Fig. 5 shows timely throughput improvement as a function of the total traffic arrival rate per region. To vary the total arrival rate, we set traffic arrival rates as $\alpha_1^1 = 0.25\gamma$ and $\alpha_1^2 = 0.75\gamma$, and vary $\gamma$ between 0 and 2. For each $T \in \{2, 4, 6, 8\}$, we see that the percentage improvement in timely throughput is low when the traffic rate is too low or too high, and it hits a peak in between. For very low traffic rates, there isn't much need for sharing in the first place and this explains the low improvements. For very high traffic rates, there isn't much room for sharing as each operator needs almost all its resources to serve its own clients.

## VI. CONCLUSIONS

Bandwidth sharing can provide significant performance improvements (up to 90%) for real-time traffic without requiring additional investment in spectrum. The performance improvements can be realized using our sharing and scheduling policy, for which we have also provided theoretical guarantees on performance. The amount of improvement depends on traffic characteristics. In particular, improvements are more when the arrival rates of different operators is imbalanced. Further, improvements are lesser when cumulative arrival rate is very low or very high.

A potential future extension of this work is a less dynamic solution for sharing, which does not require slot-by-slot coordination between operators. It will also be interesting to consider

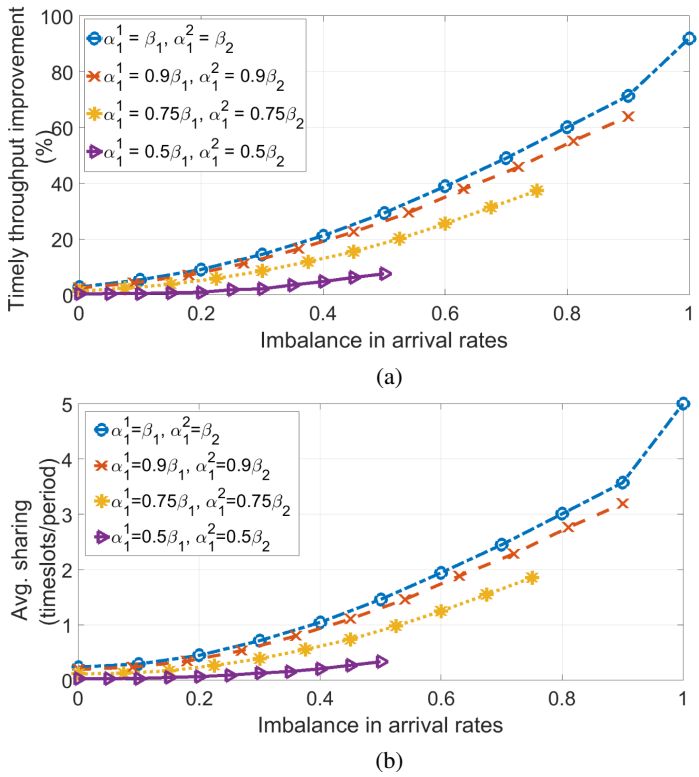

Fig. 4: Impact of imbalance in arrival rates between two operators: (a) Percentage improvement in timely throughput, (b) Average number of timeslots shared per period

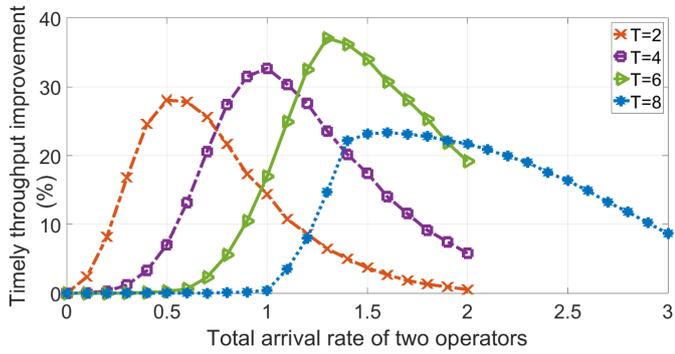

Fig. 5: Impact of total arrival rates

other (e.g., monetary) incentive mechanisms for operators to incentivize sharing.

*A. Proof of Theorem 1*

We start with some notation.

- We let $\boldsymbol{q} = (q_{r,n}^i : n \in \mathcal{N}_r^i, \forall r \in \mathcal{R}, \forall i \in \mathcal{O})$. That is, $\boldsymbol{q}$ is a multi-dimensional array with entries $q_{r,n}^i$ (defined in (1)) for each $n \in \mathcal{N}_r^i$, $\forall r \in \mathcal{R}$, $\forall i \in \mathcal{O}$.
- Let $\boldsymbol{\zeta} = (\zeta^{(i,j)} : i, j \in \mathcal{O})$. That is, $\boldsymbol{\zeta}$ is a multi-dimensional array with entries $\zeta^{(i,j)}$ (defined in (2)) for each $i, j \in \mathcal{O}$.
- Let $\boldsymbol{\delta}(k) = (\delta_{r,n}^i(k) : n \in \mathcal{N}_r^i, \forall r \in \mathcal{R}, \forall i \in \mathcal{O})$ for $k \in \mathcal{K}$. That is, $\boldsymbol{\delta}(k)$ is a multi-dimensional array with entries $\delta_{r,n}^i(k)$ (evolving according to (11)) for each $n \in \mathcal{N}_r^i$, $\forall r \in \mathcal{R}$, $\forall i \in \mathcal{O}$.
- Let $\boldsymbol{\sigma}(k) = (\sigma^{i \to j}(k) : \sigma^{i \to j}(k) : i, j \in \mathcal{O})$ for $k \in \mathcal{K}$. That is, $\boldsymbol{\sigma}(k)$ is a multi-dimensional array with entries $\sigma^{i \to j}(k)$ (updated using (12)) for each $i, j \in \mathcal{O}$.

The approach below is an extension of a proof approach in [15] which does not consider sharing.

Let $\Theta(k) = (\boldsymbol{\delta}(k), \boldsymbol{\sigma}(k))$ tracking the virtual queues under our policy evolving according to (11) and (12). Define quadratic Lyapunov function $L(\Theta(k))$ as

$$L(\Theta(k)) = \frac{1}{2}\left[\sum_{r \in \mathcal{R}} \sum_{i \in \mathcal{O}} \sum_{n \in \mathcal{N}_r^i} \delta_{r,n}^i(k)^2 + \sum_{i \in \mathcal{O}} \sum_{j \in \mathcal{O} \setminus \{i\}} \sigma^{i \to j}(k)^2\right], \forall k \in \mathcal{K}. \tag{13}$$

**Lemma 1.** *For any strictly feasible timely throughput requirement $\boldsymbol{q}$ and sharing bound $\boldsymbol{\zeta}$, there exists positive constants $C > 0$, $\alpha > 0$ and $\beta > 0$ such that for each $k \in \mathcal{K}$*

$$E\left[\frac{L(\Theta(k+K))}{K} - \frac{L(\Theta(k))}{K}\bigg|\Theta(k)\right] \leq C - \alpha \sum_{r \in \mathcal{R}} \sum_{i \in \mathcal{O}} \sum_{n \in \mathcal{N}_r^i} \delta_{r,n}^i(k) - \beta \sum_{i \in \mathcal{O}} \sum_{j \in \mathcal{O} \setminus \{i\}} \sigma^{i \to j}(k). \tag{14}$$

*Proof.* Using (11) and (12) in (13), we have

$$L(\Theta(k+1)) - L(\Theta(k)) \leq \frac{1}{2}\bigg[\sum_{r \in \mathcal{R}} \sum_{i \in \mathcal{O}} \sum_{n \in \mathcal{N}_r^i} (\delta_{r,n}^i(k) + q_{r,n}^i - b_{r,n}^{*i}(k) I_{r,n}^i(k))^2 + \tag{15}$$

$$\sum_{i \in \mathcal{O}} \sum_{j \in \mathcal{O} \setminus \{i\}} (\sigma^{i \to j}(k) + \sum_{r \in \mathcal{R}} S_r^{*j \to i}(k) - \sum_{r \in \mathcal{R}} S_r^{*i \to j}(k) - \zeta^{(i,j)})^2\bigg]$$

$$- \frac{1}{2}\bigg[\sum_{r \in \mathcal{R}} \sum_{i \in \mathcal{O}} \sum_{n \in \mathcal{N}_r^i} \delta_{r,n}^i(k)^2 + \sum_{i \in \mathcal{O}} \sum_{j \in \mathcal{O} \setminus \{i\}} \sigma^{i \to j}(k)^2\bigg]$$

$$= \frac{1}{2}\bigg[\sum_{r \in \mathcal{R}} \sum_{i \in \mathcal{O}} \sum_{n \in \mathcal{N}_r^i} (\delta_{r,n}^i(k) + q_{r,n}^i - b_{r,n}^{*i}(k) I_{r,n}^i(k))^2 + \sum_{i \in \mathcal{O}} \sum_{j \in \mathcal{O} \setminus \{i\}} (\sigma^{i \to j}(k) + \Delta S^{*(j,i)}(k) - \zeta^{(i,j)})^2\bigg]$$

$$- \frac{1}{2}\bigg[\sum_{r \in \mathcal{R}} \sum_{i \in \mathcal{O}} \sum_{n \in \mathcal{N}_r^i} \delta_{r,n}^i(k)^2 + \sum_{i \in \mathcal{O}} \sum_{j \in \mathcal{O} \setminus \{i\}} \sigma^{i \to j}(k)^2\bigg]$$

$$\tag{16}$$

$$= \frac{1}{2} \sum_{r \in \mathcal{R}} \sum_{i \in \mathcal{O}} \sum_{n \in \mathcal{N}_r^i} (q_{r,n}^i - b_{r,n}^{*i}(k) I_{r,n}^i(k))^2 + \sum_{r \in \mathcal{R}} \sum_{i \in \mathcal{O}} \sum_{n \in \mathcal{N}_r^i} \delta_{r,n}^i(k)(q_{r,n}^i - b_{r,n}^{*i}(k) I_{r,n}^i(k)) \tag{17}$$

$$+ \frac{1}{2} \sum_{i \in \mathcal{O}} \sum_{j \in \mathcal{O} \setminus \{i\}} (\Delta S^{*(j,i)}(k) - \zeta^{(i,j)})^2 + \sum_{i \in \mathcal{O}} \sum_{j \in \mathcal{O} \setminus \{i\}} \sigma^{i \to j}(k)(\Delta S^{*(j,i)}(k) - \zeta^{(i,j)}),$$

where

$$\Delta S^{*(j,i)}(k) = \sum_{r \in \mathcal{R}} S_r^{*j \to i}(k) - \sum_{r \in \mathcal{R}} S_r^{*i \to j}(k).$$

We define drift $\Delta(\Theta(k))$ in the Lyapunov function as the following conditional expectation:

$$\Delta(\Theta(k)) \triangleq E[L(\Theta(k+1)) - L(\Theta(k))|\Theta(k)]. \tag{18}$$

Using (17), we have

$$\Delta(\Theta(k)) \leq E\left[\frac{1}{2}\sum_{r\in\mathcal{R}}\sum_{i\in\mathcal{O}}\sum_{n\in\mathcal{N}_r^i}(q_{r,n}^i - b_{r,n}^{*i}(k)I_{r,n}^i(k))^2|\Theta(k)\right] + E\left[\sum_{r\in\mathcal{R}}\sum_{i\in\mathcal{O}}\sum_{n\in\mathcal{N}_r^i}\delta_{r,n}^i(k)(q_{r,n}^i - b_{r,n}^{*i}(k)I_{r,n}^i(k))|\Theta(k)\right]$$

$$+ E\left[\frac{1}{2}\sum_{i\in\mathcal{O}}\sum_{j\in\mathcal{O}\setminus\{i\}}(\Delta S^{*(j,i)}(k) - \zeta^{(i,j)})^2|\Theta(k)\right] + E\left[\sum_{i\in\mathcal{O}}\sum_{j\in\mathcal{O}\setminus\{i\}}\sigma^{i\to j}(k)(\Delta S^{*(j,i)}(k) - \zeta^{(i,j)})|\Theta(k)\right]$$

$$= B(k) + D(k) + E\left[\sum_{r\in\mathcal{R}}\sum_{i\in\mathcal{O}}\sum_{n\in\mathcal{N}_r^i}\delta_{r,n}^i(k)(q_{r,n}^i - b_{r,n}^{*i}(k)I_{r,n}^i(k))|\Theta(k)\right] +$$

$$E\left[\sum_{i\in\mathcal{O}}\sum_{j\in\mathcal{O}\setminus\{i\}}\sigma^{i\to j}(k)(\Delta S^{*(j,i)}(k) - \zeta^{(i,j)})|\Theta(k)\right] \quad (19)$$

where

$$B(k) = E\left[\frac{1}{2}\sum_{r\in\mathcal{R}}\sum_{i\in\mathcal{O}}\sum_{n\in\mathcal{N}_r^i}(q_{r,n}^i - b_{r,n}^{*i}(k)I_{r,n}^i(k))^2|\Theta(k)\right]. \quad (20)$$

$$D(k) = E\left[\frac{1}{2}\sum_{i\in\mathcal{O}}\sum_{j\in\mathcal{O}\setminus\{i\}}(\Delta S^{*(j,i)}(k) - \zeta^{(i,j)})^2|\Theta(k)\right]. \quad (21)$$

All the terms in (20) and (21) are bounded, since $|b_{r,n}^{*i}(k)I_{r,n}^i(k)| \leq 1$ and $|\Delta S^{*(j,i)}(k)| \leq 2RT$. Hence, there exist positive constants $B$ and $D$ such that $B(k) \leq B$ and $D(k) \leq D$ for each $k$. Thus, (19) implies that

$$\Delta(\Theta(k)) \leq E\left[\sum_{r\in\mathcal{R}}\sum_{i\in\mathcal{O}}\sum_{n\in\mathcal{N}_r^i}\delta_{r,n}^i(k)(q_{r,n}^i - b_{r,n}^{*i}(k)I_{r,n}^i(k))|\Theta(k)\right] + E\left[\sum_{i\in\mathcal{O}}\sum_{j\in\mathcal{O}\setminus\{i\}}\sigma^{i\to j}(k)(\Delta S^{*(j,i)}(k) - \zeta^{(i,j)})|\Theta(k)\right]$$
$$+ B + D. \quad (22)$$

Observe from IV that our sharing and scheduling policy minimizes (9) when compared to any alternative stationary randomized policy. Let $\widetilde{b}_{r,n}^i(k)$ and $\Delta\widetilde{S}^{(j,i)}(k)$ represent control actions made on period $k$ due to any other alternative stationary randomized policy. Thus, we have

$$\Delta(\Theta(k)) \leq E\left[\sum_{r\in\mathcal{R}}\sum_{i\in\mathcal{O}}\sum_{n\in\mathcal{N}_r^i}\delta_{r,n}^i(k)(q_{r,n}^i - \widetilde{b}_{r,n}^i(k)I_{r,n}^i(k))|\Theta(k)\right] + E\left[\sum_{i\in\mathcal{O}}\sum_{j\in\mathcal{O}\setminus\{i\}}\sigma^{i\to j}(k)(\Delta\widetilde{S}^{(j,i)}(k) - \zeta^{(i,j)})|\Theta(k)\right]$$
$$+ B + D. \quad (23)$$

Summing (23) yields (by noting that we have 'telescoping' sums) and dividing by $K$:

$$E\left[\frac{L(\Theta(k+K))}{K} - \frac{L(\Theta(k))}{K}\bigg|\Theta(k)\right] = E\left[\frac{\sum_{x=k}^{k+K-1}\Delta(\Theta(x))}{K}\bigg|\Theta(k)\right]$$
$$\leq \frac{1}{K}E\left[\sum_{x=k}^{k+K-1}\sum_{r\in\mathcal{R}}\sum_{i\in\mathcal{O}}\sum_{n\in\mathcal{N}_r^i}\delta_{r,n}^i(x)(q_{r,n}^i - \widetilde{b}_{r,n}^i(x)I_{r,n}^i(x))|\Theta(k)\right] +$$
$$\frac{1}{K}E\left[\sum_{x=k}^{k+K-1}\sum_{i\in\mathcal{O}}\sum_{j\in\mathcal{O}\setminus\{i\}}\sigma^{i\to j}(x)(\Delta\widetilde{S}^{(j,i)}(x) - \zeta^{(i,j)})|\Theta(k)\right] + B + D. \quad (24)$$

Simplifying the first term on the RHS,

$$\frac{1}{K}E\Bigg[\sum_{x=k}^{k+K-1}\sum_{r\in\mathcal{R}}\sum_{i\in\mathcal{O}}\sum_{n\in\mathcal{N}_r^i}\delta_{r,n}^i(x)(q_{r,n}^i-\widetilde{b}_{r,n}^i(x)I_{r,n}^i(x))|\Theta(k)\Bigg]$$

$$=\frac{1}{K}E\Bigg[\sum_{r\in\mathcal{R}}\sum_{i\in\mathcal{O}}\sum_{n\in\mathcal{N}_r^i}\sum_{x=k}^{k+K-1}\delta_{r,n}^i(x)(q_{r,n}^i-\widetilde{b}_{r,n}^i(x)I_{r,n}^i(x))|\Theta(k)\Bigg] \quad (25)$$

$$=\frac{1}{K}E\Bigg[\sum_{r\in\mathcal{R}}\sum_{i\in\mathcal{O}}\sum_{n\in\mathcal{N}_r^i}\sum_{x=k}^{k+K-1}(\delta_{r,n}^i(k)+\delta_{r,n}^i(x)-\delta_{r,n}^i(k))(q_{r,n}^i-\widetilde{b}_{r,n}^i(x)I_{r,n}^i(x))|\Theta(k)\Bigg] \quad (26)$$

$$=\frac{1}{K}E\Bigg[\sum_{r\in\mathcal{R}}\sum_{i\in\mathcal{O}}\sum_{n\in\mathcal{N}_r^i}\sum_{x=k}^{k+K-1}(\delta_{r,n}^i(x)-\delta_{r,n}^i(k))(q_{r,n}^i-\widetilde{b}_{r,n}^i(x)I_{r,n}^i(x))|\Theta(k)\Bigg]+$$

$$\frac{1}{K}E\Bigg[\sum_{r\in\mathcal{R}}\sum_{i\in\mathcal{O}}\sum_{n\in\mathcal{N}_r^i}\sum_{x=k}^{k+K-1}\delta_{r,n}^i(k)(q_{r,n}^i-\widetilde{b}_{r,n}^i(x)I_{r,n}^i(x))|\Theta(k)\Bigg] \quad (27)$$

Since $\widetilde{b}_{r,n}^i(x)$ is a non-negative bounded random variable and $q_{r,n}^i$ is also bounded, $q_{r,n}^i-\widetilde{b}_{r,n}^i(x)I_{r,n}^i(x)$ is bounded for all $n\in\mathcal{N}_r^i$, $i\in\mathcal{O}$, $r\in\mathcal{R}$ and $x\in[k,k+K-1]$. Using (11), $\delta_{r,n}^i(x)-\delta_{r,n}^i(k)=(x-k)q_{r,n}^i-\sum_{y=k}^{x-1}b_{r,n}^i(y)I_{r,n}^i(y)$. Since only $K$ consecutive intervals are considered, $\frac{\delta_{r,n}^i(x)-\delta_{r,n}^i(k)}{K}$ is also bounded for all $n\in\mathcal{N}_r^i$, $i\in\mathcal{O}$, $r\in\mathcal{R}$ and $x\in[k,k+K-1]$.

$$\frac{1}{K}E\Bigg[\sum_{x=k}^{k+K-1}\sum_{r\in\mathcal{R}}\sum_{i\in\mathcal{O}}\sum_{n\in\mathcal{N}_r^i}\delta_{r,n}^i(x)(q_{r,n}^i-\widetilde{b}_{r,n}^i(x)I_{r,n}^i(x))|\Theta(k)\Bigg]$$

$$\leq\frac{1}{K}E\Bigg[\sum_{r\in\mathcal{R}}\sum_{i\in\mathcal{O}}\sum_{n\in\mathcal{N}_r^i}\sum_{x=k}^{k+K-1}\delta_{r,n}^i(k)(q_{r,n}^i-\widetilde{b}_{r,n}^i(x)I_{r,n}^i(x))|\Theta(k)\Bigg]+C_1 \quad (28)$$

(28) can be further written as,

$$\frac{1}{K}E\Bigg[\sum_{x=k}^{k+K-1}\sum_{r\in\mathcal{R}}\sum_{i\in\mathcal{O}}\sum_{n\in\mathcal{N}_r^i}\delta_{r,n}^i(x)(q_{r,n}^i-\widetilde{b}_{r,n}^i(x)I_{r,n}^i(x))|\Theta(k)\Bigg]$$

$$\leq\frac{1}{K}E\Bigg[\sum_{r\in\mathcal{R}}\sum_{i\in\mathcal{O}}\sum_{n\in\mathcal{N}_r^i}\sum_{x=k}^{k+K-1}\delta_{r,n}^i(k)(q_{r,n}^i-\widetilde{b}_{r,n}^i(x)I_{r,n}^i(x))|\Theta(k)\Bigg]+C_1$$

$$=E\Bigg[\sum_{r\in\mathcal{R}}\sum_{i\in\mathcal{O}}\sum_{n\in\mathcal{N}_r^i}\delta_{r,n}^i(k)\bigg(q_{r,n}^i-\sum_{x=k}^{k+K-1}\frac{\widetilde{b}_{r,n}^i(x)I_{r,n}^i(x)}{K}\bigg)\bigg|\Theta(k)\Bigg]+C_1$$

$$=\sum_{r\in\mathcal{R}}\sum_{i\in\mathcal{O}}\sum_{n\in\mathcal{N}_r^i}\delta_{r,n}^i(k)E\Bigg[\bigg(q_{r,n}^i-\sum_{x=k}^{k+K-1}\frac{\widetilde{b}_{r,n}^i(x)I_{r,n}^i(x)}{K}\bigg)\bigg|\Theta(k)\Bigg]+C_1$$

$$=\sum_{r\in\mathcal{R}}\sum_{i\in\mathcal{O}}\sum_{n\in\mathcal{N}_r^i}\delta_{r,n}^i(k)\bigg(q_{r,n}^i-E\bigg[\sum_{x=k}^{k+K-1}\frac{\widetilde{b}_{r,n}^i(x)I_{r,n}^i(x)}{K}\bigg|\Theta(k)\bigg]\bigg)+C_1 \quad (29)$$

(29) was arrived at by using the fact that $\delta_{r,n}^i(k)$ is $\Theta(k)$-measurable.

Similarly, we can simplify the second term on the RHS of (24) as follows:

$$\frac{1}{K}E\left[\sum_{x=k}^{k+K-1}\sum_{i\in\mathcal{O}}\sum_{j\in\mathcal{O}\setminus\{i\}}\sigma^{i\to j}(x)(\Delta\widetilde{S}^{(j,i)}(x)-\zeta^{(i,j)})\Big|\Theta(k)\right]$$

$$=\frac{1}{K}E\left[\sum_{i\in\mathcal{O}}\sum_{j\in\mathcal{O}\setminus\{i\}}\sum_{x=k}^{k+K-1}\sigma^{i\to j}(x)(\Delta\widetilde{S}^{(j,i)}(x)-\zeta^{(i,j)})\Big|\Theta(k)\right] \quad (30)$$

$$=\frac{1}{K}E\left[\sum_{i\in\mathcal{O}}\sum_{j\in\mathcal{O}\setminus\{i\}}\sum_{x=k}^{k+K-1}(\sigma^{i\to j}(k)+\sigma^{i\to j}(x)-\sigma^{i\to j}(k))(\Delta\widetilde{S}^{(j,i)}(x)-\zeta^{(i,j)})\Big|\Theta(k)\right] \quad (31)$$

$$=\frac{1}{K}E\left[\sum_{i\in\mathcal{O}}\sum_{j\in\mathcal{O}\setminus\{i\}}\sum_{x=k}^{k+K-1}(\sigma^{i\to j}(x)-\sigma^{i\to j}(k))(\Delta\widetilde{S}^{(j,i)}(x)-\zeta^{(i,j)})\Big|\Theta(k)\right]+$$

$$\frac{1}{K}E\left[\sum_{i\in\mathcal{O}}\sum_{j\in\mathcal{O}\setminus\{i\}}\sum_{x=k}^{k+K-1}\sigma^{i\to j}(k)(\Delta\widetilde{S}^{(j,i)}(x)-\zeta^{(i,j)})\Big|\Theta(k)\right] \quad (32)$$

Since $\zeta^{(i,j)}$ is bounded and $\Delta\widetilde{S}^{(j,i)}(x)$ is bounded $\forall\ i\in\mathcal{O}, j\in\mathcal{O}\setminus\{i\}$ and $x\in[k,k+K-1]$, hence $\Delta\widetilde{S}^{(j,i)}(x)-\zeta^{(i,j)}$ is bounded $\forall\ i\in\mathcal{O}, j\in\mathcal{O}\setminus\{i\}$ and $x\in[k,k+K-1]$.

From (12), $\sigma^{i\to j}(x)-\sigma^{i\to j}(k)=\sum_{y=k}^{x-1}\Delta S^{(j,i)}(y)-(x-k)\zeta^{(i,j)}$. Since only $K$ consecutive intervals are considered, $\frac{\sigma^{i\to j}(x)-\sigma^{i\to j}(k)}{K}$ is also bounded for all $i\in\mathcal{O}, j\in\mathcal{O}\setminus\{i\}$ and $x\in[k,k+K-1]$. Hence, there exists some $C_2$ such that

$$\frac{1}{K}E\left[\sum_{x=k}^{k+K-1}\sum_{i\in\mathcal{O}}\sum_{j\in\mathcal{O}\setminus\{i\}}\sigma^{i\to j}(x)(\Delta\widetilde{S}^{(j,i)}(x)-\zeta^{(i,j)})\Big|\Theta(k)\right]$$
$$\quad (33)$$

$$\leq\sum_{i\in\mathcal{O}}\sum_{j\in\mathcal{O}\setminus\{i\}}\sigma^{i\to j}(k)\left(E\left[\sum_{x=k}^{k+K-1}\frac{\Delta\widetilde{S}^{(j,i)}(x)}{K}\Big|\Theta(k)\right]-\zeta^{(i,j)}\right)+C_2. \quad (34)$$

From the above discussion, we have

$$E\left[\frac{L(\Theta(k+K))}{K}-\frac{L(\Theta(k))}{K}\Big|\Theta(k)\right]=E\left[\frac{\sum_{x=k}^{k+K-1}\Delta(\Theta(x))}{K}\Big|\Theta(k)\right]$$

$$\leq\sum_{r\in\mathcal{R}}\sum_{i\in\mathcal{O}}\sum_{n\in\mathcal{N}_r^i}\delta_{r,n}^i(k)\left(q_{r,n}^i-E\left[\sum_{x=k}^{k+K-1}\frac{\widetilde{b}_{r,n}^i(x)I_{r,n}^i(x)}{K}\Big|\Theta(k)\right]\right)+C_1$$

$$+\sum_{i\in\mathcal{O}}\sum_{j\in\mathcal{O}\setminus\{i\}}\sigma^{i\to j}(k)\left(E\left[\sum_{x=k}^{k+K-1}\frac{\Delta\widetilde{S}^{(j,i)}(x)}{K}\Big|\Theta(k)\right]-\zeta^{(i,j)}\right)+C_2+B+D. \quad (35)$$

Consider strictly feasible timely throughput requirement $\boldsymbol{q}$ and sharing bound $\boldsymbol{\zeta}$. There exists a positive number $\kappa$ and a stationary randomized policy, which randomly decides how much an operator has to share in a period and chooses a schedule randomly, based on the packet arrivals at the beginning of this period and independent of the system history before this period, that satisfies the same constraints with timely-throughput requirement $\boldsymbol{q}/\kappa$ and sharing bound $\kappa\boldsymbol{\zeta}$. Since we model packet arrivals in each period as an irreducible finite-state Markov chain, the channel as i.i.d and $\ddot{\eta}$ is a stationary randomized policy, there exists a large enough positive number $K$ such that

$$E\left[\sum_{x=k}^{k+K-1}\frac{\ddot{b}_{r,n}^i(x)I_{r,n}^i(x)}{K}\Big|\Theta(k)\right]\geq\frac{q_{r,n}}{1-\frac{\epsilon}{2}} \quad (36)$$

$$E\left[\sum_{x=k}^{k+K-1}\frac{\Delta\ddot{S}^{(j,i)}(x)}{K}\Big|\Theta(k)\right]\leq\left(1-\frac{\epsilon}{2}\right)\zeta^{(i,j)} \quad (37)$$

where $\epsilon=1-\kappa$.

Using the above two inequalities and noting that (35) applies to the policy $\ddot{\eta}$ as well, we have

Using (36) in (29),

$$\frac{1}{K}E\left[\sum_{x=k}^{k+K-1}\sum_{r\in\mathcal{R}}\sum_{i\in\mathcal{O}}\sum_{n\in\mathcal{N}_r^i}\delta_{r,n}^i(x)(q_{r,n}^i-\ddot{b}_{r,n}^i(x)I_{r,n}^i(x))|\Theta(k)\right] \leq \sum_{r\in\mathcal{R}}\sum_{i\in\mathcal{O}}\sum_{n\in\mathcal{N}_r^i}\delta_{r,n}^i(k)\left\{q_{r,n}^i-\frac{q_{r,n}}{1-\frac{\epsilon}{2}}\right\}+C_1$$

$$=-\sum_{r\in\mathcal{R}}\sum_{i\in\mathcal{O}}\sum_{n\in\mathcal{N}_r^i}\delta_{r,n}^i(k)\left(\frac{\frac{\epsilon}{2}}{1-\frac{\epsilon}{2}}\right)q_{r,n}^i+C_1$$

$$\leq -\sum_{r\in\mathcal{R}}\sum_{i\in\mathcal{O}}\sum_{n\in\mathcal{N}_r^i}\delta_{r,n}^i(k)\left(\frac{\frac{\epsilon}{2}}{1-\frac{\epsilon}{2}}\right)q_{min}+C_1 \quad (38)$$

where $q_{min} := \min_{n\in\mathcal{N}_r^i, i\in\mathcal{O}, r\in\mathcal{R}: q_{r,n}^i > 0} q_{r,n}^i$

$q_{min}$ exists since the set over which $q_{r,n}^i$ is minimized, has finite number of elements.

Using (37) in (34),

$$\sum_{i\in\mathcal{O}}\sum_{j\in\mathcal{O}\setminus\{i\}}\sigma^{i\to j}(k)\left(E\left[\sum_{x=k}^{k+K-1}\frac{\Delta\ddot{S}^{(j,i)}(x)}{K}\Big|\Theta(k)\right]-\zeta^{(i,j)}\right)+C_2 \leq \sum_{i\in\mathcal{O}}\sum_{j\in\mathcal{O}\setminus\{i\}}\sigma^{i\to j}(k)\left\{\left(1-\frac{\epsilon}{2}\right)\zeta^{(i,j)}-\zeta^{(i,j)}\right\}+C_2$$

$$=-\sum_{i\in\mathcal{O}}\sum_{j\in\mathcal{O}\setminus\{i\}}\sigma^{i\to j}(k)\frac{\epsilon}{2}\zeta^{(i,j)}+C_2$$

$$\leq -\sum_{i\in\mathcal{O}}\sum_{j\in\mathcal{O}\setminus\{i\}}\sigma^{i\to j}(k)\frac{\epsilon}{2}\zeta_{min}+C_2 \quad (39)$$

where $\zeta_{min} := \min_{i,j\in\mathcal{O}: \zeta^{(i,j)} > 0} \zeta^{(i,j)}$

$\zeta_{min}$ exists since the set over which $\zeta^{(i,j)}$ is minimized, has finite number of elements.
Substituting (38) and (39) in (24),

$$E\left[\frac{L(\Theta(k+K))}{K}-\frac{L(\Theta(k))}{K}\Big|\Theta(k)\right] \leq C-\left(\frac{\frac{\epsilon}{2}}{1-\frac{\epsilon}{2}}\right)q_{min}\sum_{r\in\mathcal{R}}\sum_{i\in\mathcal{O}}\sum_{n\in\mathcal{N}_r^i}\delta_{r,n}^i(k)-\frac{\epsilon}{2}\zeta_{min}\sum_{i\in\mathcal{O}}\sum_{j\in\mathcal{O}\setminus\{i\}}\sigma^{i\to j}(k) \quad (40)$$

where constant $C = B+D+C_1+C_2$.

Now, (14). follows from (40) if we let $\alpha = \left(\frac{\frac{\epsilon}{2}}{1-\frac{\epsilon}{2}}\right)q_{min}$ and $\beta = \frac{\epsilon}{2}\zeta_{min}$.

$\square$

Next, we have the following result.

**Lemma 2.** *For any strictly feasible timely throughput requirement $\mathbf{q}$ and sharing bound $\boldsymbol{\zeta}$, we have*

$$\limsup_{K\to\infty}\frac{1}{K}\sum_{k=0}^{K-1}\left[E\left\{\sum_{r\in\mathcal{R}}\sum_{i\in\mathcal{O}}\sum_{n\in\mathcal{N}_r^i}\delta_{r,n}^i(kK)+\sum_{i\in\mathcal{O}}\sum_{j\in\mathcal{O}\setminus\{i\}}\sigma^{i\to j}(kK)\right\}\right] \leq \frac{C}{\alpha}+\frac{C}{\beta} \quad (41)$$

*Proof.* Replace $k$ with $kK$ in (14)

$$E\left[\frac{L(\Theta(kK+K))}{K}-\frac{L(\Theta(kK))}{K}\Big|H_{kK-1}\right] \leq C-\alpha\sum_{r\in\mathcal{R}}\sum_{i\in\mathcal{O}}\sum_{n\in\mathcal{N}_r^i}\delta_{r,n}^i(kK)-\beta\sum_{i\in\mathcal{O}}\sum_{j\in\mathcal{O}\setminus\{i\}}\sigma^{i\to j}(kK) \quad (42)$$

where $H_{kK-1}$ comprises of debts $\boldsymbol{\delta}(k)$ and $\boldsymbol{\sigma}(k)$ up to the period $(kK-1)$, i.e., the system state up to the $(kK-1)$th period. Let $\hat{L}(\Theta(k)) = L(\Theta(kK))/K$. Then, we can rewrite 42 as

$$E\left[\hat{L}(\Theta(k+1))-\hat{L}(\Theta(k))\Big|H_{kK-1}\right] \leq C-\alpha\sum_{r\in\mathcal{R}}\sum_{i\in\mathcal{O}}\sum_{n\in\mathcal{N}_r^i}\delta_{r,n}^i(kK)-\beta\sum_{i\in\mathcal{O}}\sum_{j\in\mathcal{O}\setminus\{i\}}\sigma^{i\to j}(kK). \quad (43)$$

Taking expectation on both sides

$$E\left[E\left[\hat{L}(\Theta(k+1))-\hat{L}(\Theta(k))\Big|H_{kK-1}\right]\right] \leq C-\alpha E\left[\sum_{r\in\mathcal{R}}\sum_{i\in\mathcal{O}}\sum_{n\in\mathcal{N}_r^i}\delta_{r,n}^i(kK)\right]-\beta E\left[\sum_{i\in\mathcal{O}}\sum_{j\in\mathcal{O}\setminus\{i\}}\sigma^{i\to j}(kK)\right] \quad (44)$$

Applying the law of iterated expectations on the LHS,

$$E\left[\hat{L}(\Theta(k+1))\right] - E\left[\hat{L}(\Theta(k))\right] \leq C - \alpha E\left[\sum_{r\in\mathcal{R}}\sum_{i\in\mathcal{O}}\sum_{n\in\mathcal{N}_r^i}\delta_{r,n}^i(kK)\right] - \beta E\left[\sum_{i\in\mathcal{O}}\sum_{j\in\mathcal{O}\setminus\{i\}}\sigma^{i\to j}(kK)\right]. \quad (45)$$

(45) holds for all periods. Summing over periods from 0 to $K-1$ yields a telescoping series on the LHS.

$$E\left[\hat{L}(\Theta(K))\right] - E\left[\hat{L}(\Theta(0))\right] \leq CK - \alpha\sum_{k=0}^{K-1} E\left[\sum_{r\in\mathcal{R}}\sum_{i\in\mathcal{O}}\sum_{n\in\mathcal{N}_r^i}\delta_{r,n}^i(kK)\right] - \beta\sum_{k=0}^{K-1} E\left[\sum_{i\in\mathcal{O}}\sum_{j\in\mathcal{O}\setminus\{i\}}\sigma^{i\to j}(kK)\right] \quad (46)$$

Dividing throughout by K

$$\frac{E\left[\hat{L}(\Theta(K))\right] - E\left[\hat{L}(\Theta(0))\right]}{K} \leq C - \alpha\frac{1}{K}\sum_{k=0}^{K-1} E\left[\sum_{r\in\mathcal{R}}\sum_{i\in\mathcal{O}}\sum_{n\in\mathcal{N}_r^i}\delta_{r,n}^i(kK)\right] - \beta\frac{1}{K}\sum_{k=0}^{K-1} E\left[\sum_{i\in\mathcal{O}}\sum_{j\in\mathcal{O}\setminus\{i\}}\sigma^{i\to j}(kK)\right] \quad (47)$$

Rearranging the terms,

$$\alpha\frac{1}{K}\sum_{k=0}^{K-1} E\left[\sum_{r\in\mathcal{R}}\sum_{i\in\mathcal{O}}\sum_{n\in\mathcal{N}_r^i}\delta_{r,n}^i(kK)\right] + \beta\frac{1}{K}\sum_{k=0}^{K-1} E\left[\sum_{i\in\mathcal{O}}\sum_{j\in\mathcal{O}\setminus\{i\}}\sigma^{i\to j}(kK)\right] \leq C - \frac{E[\hat{L}(\Theta(K))]}{K} + \frac{E[\hat{L}(\Theta(0))]}{K}. \quad (48)$$

Using the fact that $E[\hat{L}(\Theta(K))] \geq 0$,

$$\alpha\frac{1}{K}\sum_{k=0}^{K-1} E\left[\sum_{r\in\mathcal{R}}\sum_{i\in\mathcal{O}}\sum_{n\in\mathcal{N}_r^i}\delta_{r,n}^i(kK)\right] + \beta\frac{1}{K}\sum_{k=0}^{K-1} E\left[\sum_{i\in\mathcal{O}}\sum_{j\in\mathcal{O}\setminus\{i\}}\sigma^{i\to j}(kK)\right] \leq C + \frac{E[\hat{L}(\Theta(0))]}{K}. \quad (49)$$

(49) can be written as two inequalities

$$\alpha\frac{1}{K}\sum_{k=0}^{K-1} E\left[\sum_{r\in\mathcal{R}}\sum_{i\in\mathcal{O}}\sum_{n\in\mathcal{N}_r^i}\delta_{r,n}^i(kK)\right] \leq C + \frac{E[\hat{L}(\Theta(0))]}{K}, \quad (50)$$

$$\beta\frac{1}{K}\sum_{k=0}^{K-1} E\left[\sum_{i\in\mathcal{O}}\sum_{j\in\mathcal{O}\setminus\{i\}}\sigma^{i\to j}(kK)\right] \leq C + \frac{E[\hat{L}(\Theta(0))]}{K}. \quad (51)$$

Dividing (50) by $\alpha$ and (51) by $\beta$, and summing up, we have

$$\frac{1}{K}\sum_{k=0}^{K-1} E\left[\sum_{r\in\mathcal{R}}\sum_{i\in\mathcal{O}}\sum_{n\in\mathcal{N}_r^i}\delta_{r,n}^i(kK) + \sum_{i\in\mathcal{O}}\sum_{j\in\mathcal{O}\setminus\{i\}}\sigma^{i\to j}(kK)\right] \leq \frac{C}{\alpha} + \frac{E[\hat{L}(\Theta(0))]}{\alpha K} + \frac{C}{\beta} + \frac{E[\hat{L}(\Theta(0))]}{\beta K}.$$

Thus, we have

$$\limsup_{K\to\infty}\frac{1}{K}\sum_{k=0}^{K-1}\left[E\left[\sum_{r\in\mathcal{R}}\sum_{i\in\mathcal{O}}\sum_{n\in\mathcal{N}_r^i}\delta_{r,n}^i(kK) + \sum_{i\in\mathcal{O}}\sum_{j\in\mathcal{O}\setminus\{i\}}\sigma^{i\to j}(kK)\right]\right] \leq \frac{C}{\alpha} + \frac{C}{\beta}. \quad (52)$$

$\square$

Following is a result from [15] reproduced below for easy reference.

**Lemma 3.** *(from [15])* Let $f(n)$ be a nonnegative function such that $|f(n+1) - f(n)| \leq M$, for some $M > 0$, for all n. If $\limsup_{n\to\infty}\frac{1}{n}\sum_{i=0}^n f(i) \leq B$, for some constant B, then $\lim_{n\to\infty}\frac{1}{n}f(n) = 0$

Now, we have the following result.

**Lemma 4.** *For any strictly feasible timely throughput requirement $\mathbf{q}$ and sharing bound $\zeta$, we have*

$$Prob\left\{\frac{\delta_{r,n}^i(K)}{K} < \xi_1\right\} \to 1, \text{ as } K\to\infty, \; \forall n\in\mathcal{N}_r^i, i\in\mathcal{O}, r\in\mathcal{R}, \xi_1 > 0; \quad (53)$$

$$Prob\left\{\frac{\sigma^{i\to j}(K)}{K} < \xi_2\right\} \to 1, \text{ as } K\to\infty, \; \forall i\in\mathcal{O}, j\in\mathcal{O}\setminus\{i\}, r\in\mathcal{R}, \xi_2 > 0. \quad (54)$$

*Proof.* From Lemma 2 we have

$$\limsup_{K\to\infty} \frac{1}{K} \sum_{k=0}^{K-1} \left[ E\left[ \sum_{r\in\mathcal{R}} \sum_{i\in\mathcal{O}} \sum_{n\in\mathcal{N}_r^i} \delta_{r,n}^i(kK) + \sum_{i\in\mathcal{O}} \sum_{j\in\mathcal{O}\setminus\{i\}} \sigma^{i\to j}(kK) \right] \right] \leq \frac{C}{\alpha} + \frac{C}{\beta}.$$

Note that

$$\left| \sum_{r\in\mathcal{R}} \sum_{i\in\mathcal{O}} \sum_{n\in\mathcal{N}_r^i} \delta_{r,n}^i(kK+K) + \sum_{i\in\mathcal{O}} \sum_{j\in\mathcal{O}\setminus\{i\}} \sigma^{i\to j}(kK+K) - \sum_{r\in\mathcal{R}} \sum_{i\in\mathcal{O}} \sum_{n\in\mathcal{N}_r^i} \delta_{r,n}^i(kK) - \sum_{i\in\mathcal{O}} \sum_{j\in\mathcal{O}\setminus\{i\}} \sigma^{i\to j}(kK) \right|$$

is bounded, since for each $k$

$$\left| \sum_{r\in\mathcal{R}} \sum_{i\in\mathcal{O}} \sum_{n\in\mathcal{N}_r^i} \delta_{r,n}^i(kK+K) - \sum_{r\in\mathcal{R}} \sum_{i\in\mathcal{O}} \sum_{n\in\mathcal{N}_r^i} \delta_{r,n}^i(kK) \right| \text{ is bounded,}$$

$$\left| \sum_{i\in\mathcal{O}} \sum_{j\in\mathcal{O}\setminus\{i\}} \sigma^{i\to j}(kK+K) - \sum_{i\in\mathcal{O}} \sum_{j\in\mathcal{O}\setminus\{i\}} \sigma^{i\to j}(kK) \right| \text{ is bounded.}$$

Now using Lemma 3 (from [15]), we have

$$\lim_{K\to\infty} \frac{1}{K} E\left[ \sum_{r\in\mathcal{R}} \sum_{i\in\mathcal{O}} \sum_{n\in\mathcal{N}_r^i} \delta_{r,n}^i(kK) + \sum_{i\in\mathcal{O}} \sum_{j\in\mathcal{O}\setminus\{i\}} \sigma^{i\to j}(kK) \right] = 0. \tag{55}$$

Observe that (55) signifies convergence in mean, which also implies convergence in probability i.e.,

$$\frac{1}{K} \left[ \sum_{r\in\mathcal{R}} \sum_{i\in\mathcal{O}} \sum_{n\in\mathcal{N}_r^i} \delta_{r,n}^i(kK) + \sum_{i\in\mathcal{O}} \sum_{j\in\mathcal{O}\setminus\{i\}} \sigma^{i\to j}(kK) \right] \tag{56}$$

converges to 0 in probability.

$$|\delta_{r,n}^i(kK) + \sigma^{i\to j}(kK) - \delta_{r,n}^i(x) - \sigma^{i\to j}(x)| \text{ is bounded } \forall x \in [kK, kK+K] \text{ and } K \text{ is also bounded}$$

So, we have

$$\frac{\delta_{r,n}^i(K)}{K} \text{ converges to 0 in probability } \forall n \in \mathcal{N}_r^i, i \in \mathcal{O}, r \in \mathcal{R}$$

and

$$\frac{\sigma^{i\to j}(K)}{K} \text{ converges to 0 in probability } \forall i \in \mathcal{O}, j \in \mathcal{O}\setminus\{i\}$$

That is

$$\text{Prob}\left\{ \frac{\delta_{r,n}^i(K)}{K} < \xi_1 \right\} \to 1, \text{ as } K \to \infty, \ \forall n \in \mathcal{N}_r^i, i \in \mathcal{O}, r \in \mathcal{R}, \xi_1 > 0;$$

$$\text{Prob}\left\{ \frac{\sigma^{i\to j}(K)}{K} < \xi_2 \right\} \to 1, \text{ as } K \to \infty, \ \forall i \in \mathcal{O}, j \in \mathcal{O}\setminus\{i\}, \xi_2 > 0.$$

□

**Lemma 5.** *For any strictly feasible timely throughput requirement $\mathbf{q}$ and sharing bound $\boldsymbol{\zeta}$, (1) and (2) are satisfied.*

*Proof.* From Lemma 4 we obtain

$$\text{Prob}\left\{ \frac{\delta_{r,n}^i(K)}{K} < \xi_1 \right\} \to 1, \text{ as } K \to \infty, \ \forall n \in \mathcal{N}_r^i, i \in \mathcal{O}, r \in \mathcal{R}, \xi_1 > 0;$$

$$\text{Prob}\left\{ \frac{\sigma^{i\to j}(K)}{K} < \xi_2 \right\} \to 1, \text{ as } K \to \infty, \ \forall i \in \mathcal{O}, j \in \mathcal{O}\setminus\{i\}, \xi_2 > 0.$$

Consider the update equation of $\sigma^{i\to j}(k)$ given in (12). It can be re-written as

$$\sigma^{i\to j}(k+1) \geq \sigma^{i\to j}(k) + \sum_{r\in\mathcal{R}} S_r^{j\to i}(k) - \sum_{r\in\mathcal{R}} S_r^{i\to j}(k) - \zeta, \quad \forall i \in \mathcal{O}, j \in \mathcal{O}\setminus\{i\}, k \in \mathcal{K}. \tag{57}$$

Average the above equation over all periods

$$\frac{\sum_{k=0}^{K-1} \sigma^{i \to j}(k+1)}{K} \geq \frac{\sum_{k=0}^{K-1} \sigma^{i \to j}(k)}{K} + \frac{1}{K}\left[\sum_{r \in \mathcal{R}} S_r^{j \to i}(k) - \sum_{r \in \mathcal{R}} S_r^{i \to j}(k)\right] - \zeta, \quad \forall i \in \mathcal{O}, j \in \mathcal{O} \setminus \{i\}, k \in \mathcal{K}. \tag{58}$$

$$\frac{\sum_{k=0}^{K-1} \sigma^{i \to j}(k+1)}{K} - \frac{\sum_{k=0}^{K-1} \sigma^{i \to j}(k)}{K} \geq \frac{1}{K}\left[\sum_{r \in \mathcal{R}} S_r^{j \to i}(k) - \sum_{r \in \mathcal{R}} S_r^{i \to j}(k)\right] - \zeta, \quad \forall i \in \mathcal{O}, j \in \mathcal{O} \setminus \{i\}, k \in \mathcal{K}. \tag{59}$$

LHS yields a telescoping series. Applying the initial condition of $\sigma^{i \to j}(k)$

$$\frac{\sigma^{i \to j}(K)}{K} \geq \frac{1}{K}\left[\sum_{r \in \mathcal{R}} S_r^{j \to i}(k) - \sum_{r \in \mathcal{R}} S_r^{i \to j}(k)\right] - \zeta, \quad \forall i \in \mathcal{O}, j \in \mathcal{O} \setminus \{i\}, k \in \mathcal{K}. \tag{60}$$

Applying probability on both sides

$$\Pr\left\{\frac{\sigma^{i \to j}(K)}{K}\right\} \geq \Pr\left\{\frac{1}{K}\left[\sum_{r \in \mathcal{R}} S_r^{j \to i}(k) - \sum_{r \in \mathcal{R}} S_r^{i \to j}(k)\right]\right\} - \zeta, \quad \forall i \in \mathcal{O}, j \in \mathcal{O} \setminus \{i\}, k \in \mathcal{K}. \tag{61}$$

Using Lemma 4 in (61), we obtain

$$\text{Prob}\left\{\frac{1}{K}\left|\sum_{k=1}^{K}\sum_{r \in \mathcal{R}} S_r^{j \to i}(k) - \sum_{k=1}^{K}\sum_{r \in \mathcal{R}} S_r^{i \to j}(k)\right| \leq \zeta + \xi_2\right\} \to 1 \text{ as } K \to \infty,$$

□